\documentclass[twocolumn,pre,nofootinbib,floatfix,superscriptaddress]{revtex4}

\usepackage{amsmath,amsfonts,amssymb}
\usepackage{graphicx}

\newcommand{\e}{\mathrm{e}}
\newcommand{\eq}{\mathrm{*}}
\renewcommand{\d}{\mathrm{d}}

\begin{document}

\title{A generalized aggregation-disintegration model \\ for the frequency of severe terrorist attacks\footnote{The journal version of this pre-print appeared as 
{\em Journal of Conßict Resolution}, {\bf 54}(1): 179--197 (2010), which can be found at 
{\tt http://jcr.sagepub.com/cgi/content/abstract/54/1/179} . \vspace{1mm}\\ 
Address for correspondence: Aaron Clauset, 1399 Hyde Park Rd., Santa Fe, NM, 87501, USA. \vspace{1mm}\\ E-mail: {\tt aaronc@santafe.edu}, {\tt f.w.wgl@hotmail.com} } }
\author{Aaron Clauset}
\affiliation{Santa Fe Institute, New Mexico}
\author{Frederik W. Wiegel}
\affiliation{Institute of Theoretical Physics, University of Amsterdam, The Netherlands}

\begin{abstract}
We present and analyze a model of the frequency of severe terrorist attacks, which generalizes the recently  proposed model of Johnson et al. This model, which is based on the notion of self-organized criticality and which describes how terrorist cells might aggregate and disintegrate over time, predicts that the distribution of attack severities should follow a power-law form with an exponent of $\alpha=5/2$. This prediction is in good agreement with current empirical estimates for terrorist attacks worldwide, which give $\hat{\alpha}=2.4\pm0.2$, and which we show is independent of certain details of the model. We close by discussing the utility of this model for understanding terrorism and the behavior of terrorist organizations, and mention several productive ways it could be extended mathematically or tested empirically. \\

\end{abstract}

\maketitle

Richardson's Law -- one of the few robust statistical regularities in studies of political conflict -- states that the distribution of casualties in violent conflicts follows a power-law form, in which the probability of an event with $x$ deaths is \mbox{$p(x) \propto x^{-\alpha}$} where $\alpha$ is a parameter called the scaling exponent~\cite{richardson:1948,richardson:1960,cederman:2003}. Recent studies have used rigorous statistical methods to confirm this statistical law for wars between 1816 and 1980~\cite{newman:2005,clauset:etal:2009}, and have extended it to cover the severity of individual terrorist attacks, worldwide from 1968 to 2008~\cite{clauset:young:gleditsch:2007,clauset:etal:2009}.

Although Richardson's original interest in violent conflicts were inclusive of both wars and homicides, most research on the severity of conflicts has focused on wars and other large-scale events, characterizing them mainly dichotomously according to their incidence or absence (with some exceptions; see Cederman~\cite{cederman:2003} and Lacina~\cite{lacina:2006}). Research on terrorism has tended to be similarly focused~\cite{li:2005}, with considerable additional attention paid to its strategic elements~\cite{enders:sandler:2004,pape:2003}. As a result, little is known systematically about what factors and mechanisms influence the severity of terrorist attacks. This ignorance is exacerbated in part by the extreme scarcity of systematic, quantitative data on, for instance, the recruitment, fundraising, decision making, and structure of terrorist organizations, or on the counter-terrorism efforts of states. But, good-quality data on terrorist attacks themselves do exist, and their systematic analysis led to the discovery that Richard's Law includes the severity of terrorist attacks.

Power-law distributions have recently attracted a great deal of interest across the sciences, and have been found to characterize the distribution of a wide variety of natural and social phenomena. Examples of power-law distributed quantities include earthquakes, floods and forest fires~\cite{bak:tang:wiesenfeld:1987,malamud:etal:1998,newman:2005}, as well as city sizes, citation counts for scientific papers, the number of participants in strikes, and the frequency of words in written language~\cite{zipf:1949,simon:1955,newman:2005,biggs:2005}. These distributions are scientifically interesting because they depart dramatically from Central Limit Theorem assumptions of normality (or even log-normality), and extremely large or severe events are orders of magnitude more likely than would normally be expected. Further, the discovery that an empirical quantity follows a power-law distribution suggests certain unusual kinds of mechanistic explanations for their origin, e.g., mechanisms that rely on long-range correlations, long-term memory effects, or positive feedbacks. In the case of terrorism, an understanding of the social or political mechanisms the govern the frequency of severe terrorist attacks would have strong implications for security policies. (Readers unfamiliar with power-law distributions can refer to Appendix~\ref{appendix:powerlaws} for a brief primer, or to reviews by Kleiber and Kotz~\cite{kleiber:kotz:2003}, Newman~\cite{newman:2005} and Mitzenmacher~\cite{mitzenmacher:2004:b}.)

\begin{figure}[t]
\begin{center}
\includegraphics[scale=0.58]{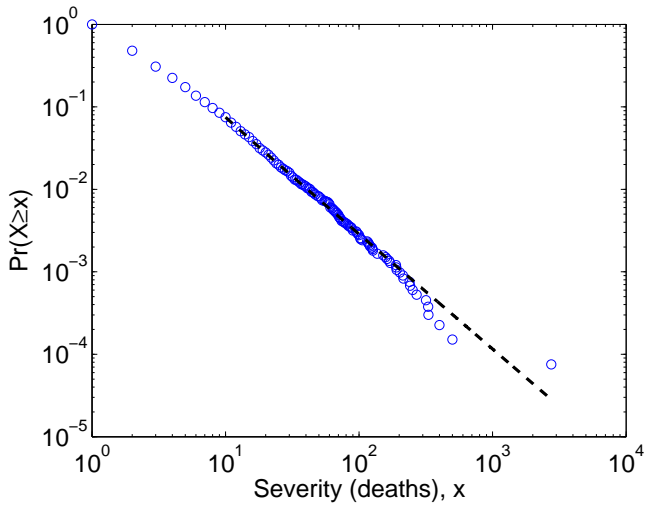}
\end{center}
\caption{The severities (number of deaths) for 13,274 fatal terrorist attacks worldwide from 1968--2008~\cite{mipt:2008}. The data are plotted as a complementary cumulative distribution function $\Pr(X\geq x)$. The solid black line shows the power-law behavior of the distribution, with scaling exponent $\hat{\alpha}=2.4\pm0.2$ for $x\geq10$~\cite{clauset:etal:2009}.}
\label{fig:severities}
\end{figure}

For terrorist attacks, recent analyses of empirical data suggest that the distribution of event severities, i.e., the number of deaths or casualties, follows such a power law, and that this statistical pattern has been largely stable over the past 40 years despite large changes in the global political system over the same period~\cite{clauset:young:gleditsch:2007}. (For concreteness, we reproduce this result from Clauset, Young, and Gleditsch in Fig.~\ref{fig:severities}.) Clauset, Young, and Gleditsch further showed that the severity of terrorist attacks remains power-law distributed, although with different scaling exponents, even after controlling for the type of weapon used (e.g., firearms, explosives, etc.) or the level of economic development of the target country, but not when controlling for geographic region (e.g., North America, Europe, etc.) or tactic (e.g., hostage, assassination, suicide bombing). Other studies of the severity of such attacks go further, suggesting that the frequency and severity of events within individual conflicts, such as those in Colombia and Iraq, exhibit power-law statistics~\cite{johnson:etal:2005,johnson:etal:2006,cioffi-revilla:romero:2009}, and that observable changes in the power law's exponent over time are indicative of real and important shifts in the underlying dynamics of the social and political generative processes.

At present, these ubiquitous power-law statistics lack a clear and well-supported explanation: what mechanisms, political or otherwise, give rise to these law-like behaviors? A scientific answer to this question may ultimately shed light, in a manner complementary to traditional studies, on the use of such tactics in violent conflicts~\cite{pape:2003,li:2005}, the internal dynamics of terrorist organizations~\cite{cordes:etal:1985,clauset:gleditsch:2009}, and trends in global terrorism~\cite{enders:sandler:2000,state:2004}. It may also shed light on the connection between severity and other modalities~\cite{clauset:etal:2009:b}, e.g., location and timing, suggest novel intervention strategies or policy recommendations for counter-terrorism~\cite{cronin:2003}, and shed light on the connection between terrorism and other kinds of violent conflicts, such as civil and international wars~\cite{richardson:1960,small:singer:1982}.

To date, two explanations have been proposed for the origin of the observed power law in the frequency of severe terrorist attacks.\footnote{We note that a wide variety of mechanisms can produce power-law distributions. Most of these processes, however, are not well-suited for explaining the severity of terrorist attacks (see Clauset, Young, and Gleditsch~\cite{clauset:young:gleditsch:2007} for some discussion). As such, we focus our attention on the two mechanisms that have been proposed, both of which have some empirical support.} One, proposed by Clauset, Young, and Gleditsch~\cite{clauset:young:gleditsch:2007} relies on an exponential sampling mechanism in which states and terrorists compete to decide which planned events become real. In this model, terrorists invest time planning events and the potential severity of these increases roughly exponentially with the total planning time. Through counter-terrorism actions by states, along with other natural attrition factors, these potential events are then strongly sampled, with the probability that a potential event becomes real decreasing roughly exponentially with the size of the event. That is, large events are exponentially less likely to become real than smaller events. The competition of these two exponentials produces a power-law distribution in the severity of events, with the scaling exponent $\alpha$ depending only on the two exponential rates.

The second mechanism, proposed by Johnson et al.~\cite{johnson:etal:2005,johnson:etal:2006}, is a self-organized critical model~\cite{bak:tang:wiesenfeld:1987} of the internal dynamics of a modern terrorist organization. In this model, a terrorist organization is composed of cells that merge and fall apart according to simple probabilistic rules (see below). The long-term dynamics of this aggregation-disintegration process produces a dynamic equilibrium or steady-state that is characterized by a power-law distribution in the sizes of cells, and, by assumption, a power-law distribution in the severity of events. In this model the scaling exponent in the steady-state can be calculated exactly, and is found to be $\alpha=5/2$. This value is in good agreement with the best current empirical estimate of \mbox{$\hat{\alpha}=2.4\pm0.2$}~\cite{clauset:etal:2009} for terrorist attacks worldwide from 1968 to 2008.

In this article, we mathematically study the Johnson et al.\ model. In particular, we generalize Johnson et al.'s specific model to a family of such models. We then analytically solve for their steady-state behavior, and show that a power-law distribution is a {\em universal} feature\footnote{Here, \emph{universality} denotes the robustness of certain qualitative features of a mathematical model to certain specific modeling assumptions. This usage is distinct from, and should not be confused with, the less technical usage of the same term to denote a natural or social phenomenon that appears to be independent of certain contingent or contextual details.} of this class of models. That is, provided the number $N$ of radicalized individuals is large $N\gg1$, the appearance of the power-law distribution and the value of its scaling exponent $\alpha$ does not depend on certain details of the model itself. Mathematically speaking: our analysis is exact in the limit $N\rightarrow\infty$. We note that our asymptotic analysis is done purely for mathematical convenience; the limit $N\rightarrow\infty$ has no social meaning and so long as $N$ is very large, our results should hold.

The benefits of generalizing the Johnson et al.\ model are two fold. First, there is the generalization itself, which extends the model in a new and important direction, and demonstrates that the model's main qualitative result---the power-law distribution in event sizes---is robust to certain specific modeling assumptions. Second, by carefully describing the model's assumptions and then mathematically working out their consequences, we can more precisely identify which empirical tests are ultimately necessary to support or refute the model's assumptions and predictions. This approach defers answering the question of what mechanism produces the power-law distribution in the frequency of severe terrorist attacks; however, this seems acceptable partly because of the complexity of the model and its analysis, and partly because of our currently very limited knowledge of the social and political processes that might give rise to the power-law distribution. This model-based approach can thus highlight which empirical facts it would be useful to know and stimulate research in productive directions.

\section{The model}
\label{sec:model}
The model we analyze is based on five assumptions about the interaction of the terrorist cells that make up a modern terrorist organization. We make no other assumptions about the relationship between these cells and the conflict or the terrorist organization they inhabit, the mode of attack or tactic used by an attacking cell, or that this model represents the behavior of hierarchical terrorist organizations.

Although these assumptions are straightforward to state, and allow us to mathematically analyze their consequences, they embody strong and possibly unrealistic constraints on the internal dynamics of terrorist groups that have not yet been systematically tested with empirical data. At present, however, this model is worthwhile to study mainly because it yields one prediction---a power-law distribution in the frequency and severity of events---that agrees relatively well with a wide range of empirical data~\cite{johnson:etal:2005,johnson:etal:2006,clauset:young:gleditsch:2007,cioffi-revilla:romero:2009,clauset:etal:2009}. By carefully exploring the behavior of this model, we can identify quantitative predictions or critical assumptions that may be tested using the available empirical data. In our concluding remarks, we discuss some of these tests and possible extensions of the model that relax some of the model's assumptions.

The five model assumptions are
\begin{enumerate}
\item There is a ``pool'' of $N$ radicalized individuals that are ``inclined'' toward terrorism. We assume $N$ to be large $N\gg1$ and to be constant in time. This latter assumption implies that terrorists who are eliminated for any reason, e.g., by counter-terrorism measures, inter- or intra-cell conflict, personal preferences, or in the course of their attacks, are replaced immediately by an equal number of radicalized individuals.
\item These individuals can form cells of size $1,2,3,4,\dots$ Let $n_{k}$ denote the number of cells consisting of \mbox{$k=1,2,3,\dots$} individuals.
\item Cells grow by a process of aggregation, in which any pair of cells can merge to form a larger cell. Specifically, we assume that any pair of cells consisting of $k$ and $\ell$ individuals respectively has a probability $A_{0}(k\,\ell)^{a}$ per unit time to combine into a cell of size $k+\ell$. Here $A_{0}>0$ and $a\geq0$ are parameters of the model, and we analyze the model for general $a$. To be realistic when comparing with data, however, we choose $a\cong1$ to represent the fact that the number of possible human relations between members of the two cells is $k\,\ell$, i.e., it scales linearly with the product of the cell sizes.
\item Cells fall apart or ``disintegrate'' spontaneously into single individuals. Let $b(k)$ denote the probability per unit time that a given cell of $k$ individuals will disintegrate spontaneously into $k$ cells of size one, and where $b(1)=0$. The explicit form of the function $b(k)$ is not needed to calculate the equilibrium distribution of cell size, provided one studies the asymptotic region $N\gg1$.
\item At any time, any cell can launch an attack. For simplicity, we assume that the attack occurs with probability (per unit time) that is independent of the cell's size, its ``age'', the number of attacks it has previously launched, etc., and that the severity $v(k)$ of an attack is roughly proportional to the cell's size $k$, i.e., $v(k) \propto k$, for $1\ll k\ll N$.
\end{enumerate}
To be precise, the number of possible pairings of a $k$-cell with a $\ell$-cell, i.e., the number of potential combinations between some cell of size $k$ and some cell of size $\ell$, equals $n_{k}n_{\ell}$ for $k\not=\ell$, and $\frac{1}{2}n_{k}(n_{k}-1)$ for $k=\ell$. However, if $N\gg1$, we shall find that all $n_{k}\gg1$; in this case we can approximate $\frac{1}{2}n_{k}(n_{k}-1) \cong \frac{1}{2}n_{k}^{2}$, which simplifies the mathematics considerably but does not fundamentally alter the results.

Our analysis of this model will show that the steady-state distribution of the sizes of the terrorist cells follows a power-law distribution with exponent $\alpha=5/2$. By assumption 5, that the severity of an attack is proportional to the size of the attacking cell, this then implies that the distribution of event severities follows a power-law distribution with the same exponent.

\section{The distribution of cell sizes in the steady state}
\label{sec:distribution}
From the five assumptions discussed above, we can write down the equation for how $n_{k}(t)$ changes with time for $k=2,3,\dots$
\begin{align}
\label{eq:3.1}
\frac{\d n_{k}}{\d t} = &  \frac{1}{2} A_{0} \sum_{i,j=1}^{\infty}\!\!{}^{'}\, i^{a} j^{a} n_{i} n_{j}  - A_{0} k^{a} n_{k} \sum_{j=1}^{\infty} j^{a} n_{j} - b(k) \, n_{k} \enspace ,
\end{align}
where $\sum\!{}^{'}\,$ denotes a summation over all natural numbers $i$ and $j$ such that
\begin{align}
i+j=k\enspace .
\end{align}
The equation for $\d n_{1} / \d t$ is not needed in our analysis. In words, the first term on the right-hand side of Eq.~\eqref{eq:3.1} represents the increase of the number of cells of size $k$ because of the aggregation of two smaller cells, the second term measures the decrease of this number because such a cell can itself merge with another cell, and the third term represents the loss of these cells because of spontaneous disintegration.

As we are interested mainly in the steady-state behavior of this model, we denote $\lim_{t\rightarrow\infty} n_{k}(t)$ by $n_{k}^{\eq}$, where ${}^{\eq}$ is not an exponent but a label that denotes the attached variable being in its steady-state limit. Eq.~\eqref{eq:3.1} now simplifies to 
\begin{align}
\label{eq:3.3}
\frac{1}{2} A_{0} & \sum_{i,j}\!{}^{'}\, i^{a}j^{a}n_{i}^{\eq}n_{j}^{\eq} = A_{0}k^{a}n_{k}^{\eq}\sum_{j} j^{a}n_{j}^{\eq} + b(k)\,n_{k}^{\eq} \enspace ,
\end{align}
for $k=2,3,\dots$. As a technical detail, we point out that the term with $j=k$ in the second summation in the right-hand sides of Eqs.~\eqref{eq:3.1} and~\eqref{eq:3.3} comes from the fact that the number of pairs $k,k$ equals $\frac{1}{2}n_{k}^{2}$ (see Section~\ref{sec:model}), but as each combination of two such cells leads to the decrease of $n_{k}$ by two, the loss term is proportional to $2\cdot\frac{1}{2}n_{k}^{2}=n_{k}^{2}$.

A simple way of solving the set of equations given in Eq.~\eqref{eq:3.3} is by introducing the generating functions~\cite{wilf:2006}
\begin{align}
\label{eq:3.4}
f(z) & \equiv \sum_{k=1}^{\infty} k^{a}n_{k}^{\eq}z^{k} \\
g(z) & \equiv \sum_{k=1}^{\infty} b(k)\,n_{k}^{\eq}z^{k} \enspace .
\end{align}
That is, we multiply Eq.~\eqref{eq:3.3} by $z^{k}$ and then sum over $k$ from $2$ to $\infty$. This reduces our system of equations to
\begin{align}
\label{eq:3.6}
\frac{1}{2}A_{0}\,f(z)\,f(z) = A_{0}\,f(1)\left\{f(z) - n_{1}^{\eq}z\right\} + g(z) \enspace ,
\end{align}
where we used the fact that $b(1)=0$ because a cell of one individual cannot disintegrate into single individuals. (Readers unfamiliar with generating functions can refer to Appendix~\ref{appendix:gfun} for a brief primer, and to Wilf~\cite{wilf:2006} for a more thorough introduction.)

Although the solution of Eq.~\eqref{eq:3.6} is difficult for general $z$ and $N$, it is much simpler in our case where $z$ is fixed and the limit $N\rightarrow\infty$ is studied. For $N\gg1$, the equilibrium frequencies $n_{k}^{\eq}$ will be proportional to $N$ (for $k$ smaller than some cut-off $k_{0}$ which we need not calculate explicitly; see Appendix~\ref{appendix:cutoff}). Hence the leading orders of magnitude (in $N$) of the various terms in Eq.~\eqref{eq:3.6} are
\begin{align}
f(z) & \sim N \\
g(z) & \sim N \\
\frac{1}{2}A_{0}\,f(z)\,f(z) & \sim N^{2} \\
A_{0}\,f(1)\left\{f(z)-n_{1}^{\eq}z\right\} & \sim N^{2} \enspace .
\end{align}
This means that for $z$ fixed and $N\gg1$, Eq.~\eqref{eq:3.6} can be replaced by
\begin{align}
\frac{1}{2}f^{2}(z)-f(1)\,f(z)+f(1)\,n_{1}^{\eq}z = 0 \enspace ,
\end{align}
which has the solution
\begin{align}
f(z) = f(1) - \sqrt{f^{2}(1) - 2f(1)\,n_{1}^{\eq}z} \enspace .
\end{align}
Substituting $z=1$ shows
\begin{align}
\label{eq:15}
f(1) = 2n_{1}^{\eq} \enspace ,
\end{align}
and gives
\begin{align}
\label{eq:3.11}
f(z) = 2n_{1}^{\eq}\left\{1-\sqrt{1-z}\right\} \enspace .
\end{align}

The definition of $f(z)$ given in Eq.~\eqref{eq:3.4} shows that the term $k^{a}n_{k}^{\eq}$ can now be found as the coefficient of $z^{k}$ in the power series expansion of Eq.~\eqref{eq:3.11}. For small values of $k$ these coefficients can be calculated by hand from the series
\begin{align}
f(z) = & 2n_{1}^{\eq}\left(\frac{1}{2}z  \,\,+\,\, \frac{1}{2}\cdot\frac{1}{4}z^{2}\,\,+\,\,\frac{1}{2}\cdot\frac{1}{4}\cdot\frac{3}{6}z^{3} \right. \nonumber \\
 & \hspace{1cm} \left. \,\,+\,\,  \frac{1}{2}\cdot\frac{1}{4}\cdot\frac{3}{6}\cdot\frac{5}{8}z^{4} \,\,+\,\, \dots \right) \enspace .
\end{align}
For example, the first four terms are
\begin{align}
2^{a}n_{2}^{\eq} & = \frac{1}{4}n_{1}^{\eq} \enspace ,\\
3^{a}n_{3}^{\eq} & = \frac{1}{8}n_{1}^{\eq} \enspace ,\\
4^{a}n_{4}^{\eq} & = \frac{5}{64}n_{1}^{\eq} \enspace ,\\
\label{eq:3.13d} 5^{a}n_{5}^{\eq} & = \frac{7}{128}n_{1}^{\eq} \enspace .
\end{align}
To obtain the coefficients for $k\gg1$, one can use Cauchy's theorem, which gives the contour integral
\begin{align}
k^{a}n_{k}^{\eq} = \imath\, \frac{n_{1}^{\eq}}{\pi} \oint_{C}z^{-k-1}\sqrt{1-z}\,\, \d z \enspace ,
\end{align}
where the contour $C$ encircles the origin of the complex $z$-plane once in the counter-clockwise direction. This contour can be deformed into a contour $C'$ which encircles the branch cut $1\leq z < \infty$ once in clockwise direction. For $z$ near to the branch point at $z=1$, it is convenient to first write
\begin{align}
z & = 1 + \zeta \\
z^{-k-1} & \cong \e^{-(k+1)\zeta} \enspace .
\end{align}
When $\zeta$ has a small positive imaginary part, one can write $\sqrt{-\zeta} = -\imath\sqrt{|\zeta|}$; when $\zeta$ has a small negative imaginary part, one writes $\sqrt{-\zeta} = +\imath\sqrt{|\zeta|}$. Hence we find the asymptotic result
\begin{align}
k^{a}n_{k}^{\eq} & \cong \frac{2}{\pi} n_{1}^{\eq} \int_{0}^{\infty} \sqrt{\zeta}\,\e^{-(k+1)\zeta}\,\d \zeta \nonumber \\
& = \frac{1}{\sqrt{\pi}}n_{1}^{\eq}(k+1)^{-3/2} \enspace ,
\end{align}
for $k\gg1$. (An alternative approach to this result would express $\{1-\sqrt{1-z}\}$ as a ratio of $\Gamma$-functions and use asymptotic analysis.) For $k$ as small as $5$, the last equation gives reasonably close approximations of the true values, e.g., for $5^{a}n_{5}^{\eq}$ the value of $0.038$, where as the exact value [from Eq.~\eqref{eq:3.13d}] is $0.055$.

This analysis thus shows that the number of cells consisting of $k$ terrorists, at equilibrium, is given by the power law
\begin{align}
\label{eq:26}
n_{k}^{\eq}\cong \frac{1}{\sqrt{\pi}}n_{1}^{\eq}k^{-a-3/2} \enspace ,
\end{align}
for $k\gg1$. Hence, because of model assumption 5, that the severity of an event is proportional to the size of the attacking cell, the probability $p_{k}$ that a terrorist attack will claim $k$ victims will also have a power-law distribution is
\begin{align}
\label{eq:3.19}
p_{k}\propto k^{-\alpha} \enspace ,
\end{align}
for $k\gg1$, with an exponent
\begin{align}
\alpha = a+3/2 \enspace .
\end{align}
As mentioned before, we assume that $a\cong1$ (see Section~\ref{sec:model}), which leads to the prediction 
\begin{align}
\alpha = 5/2 \enspace .
\label{eq:5:2}
\end{align}
In fact, for $a=1$ and $b(k)\propto k$, this model can be solved exactly, i.e., with no approximations, and doing so recovers the results of Johnson et al.~\cite{johnson:etal:2005,johnson:etal:2006}.

The value in Eq.~\eqref{eq:5:2} is in good agreement with recent estimates from empirical data~\cite{clauset:young:gleditsch:2007,clauset:etal:2009}, which give $\hat{\alpha} = 2.4\pm0.2$ for terrorist attacks worldwide since 1968.

\section{Concluding remarks}
Thus we find that the class of dynamical models studied here produces a steady state in which the number of terrorist cells of size $k$, and by assumption the severity of their attacks, follows a power-law distribution. This feature implies that the dynamics of this model system are characterized by self-organized criticality~\cite{bak:tang:wiesenfeld:1987}. Further, we find that the scaling exponent of this distribution $\alpha = 5/2$ is (for $N\gg1$) independent of the manner in which terrorist cells disintegrate [represented by the function $b(k)$]. That is, whether cells tend to disintegrate due to internal conflict, external efforts, some combination of these or other factors does not change the fundamental character of the frequency-severity distribution of attacks. In this sense, the statistical properties predicted by these models show a form of universality.

However, other statistical properties of the model should depend on the function $b(k)$ in a crucial way. For the record, we give three such properties.
\begin{itemize}
\item The explicit determination of $n_{1}^{\eq}$, the number of lone terrorists, as a function of $N$, the number of radicalized individuals.
\item A terrorist cell will grow in the course of time by combining occasionally with a smaller cell. As a result, the size of a particular cell will be time-dependent. For $a=1$ in particular, we find that the size of a terrorist cell increases exponentially with time. Similarly, each cell of size $k>2$ has a probability to disintegrate, which will also be time-dependent. 
\item The previous problem is especially interesting if one starts with a single, radicalized individual. The theory presented here makes it possible to calculate the ``speed'' with which such an individual cycles through cells of various sizes, in the steady state. 
\end{itemize}

From a policy perspective, an important question for this model concerns the difficulty of inducing qualitative changes in the steady-state behavior via realistic interventions. For instance, the independence of the model system's behavior from the particular manner in which cells disintegrate suggests that efforts focused mainly on breaking-up terrorist cells may not produce long-term changes in the severity of terrorist attacks unless they are paired with additional interventions, such as reducing the pool of radicalized individuals by other means. On the other hand, the aggregation process, i.e., the manner in which terrorist cells can achieve coordinated behavior, is a clear target, and its frustration may have a strong influence on the frequency of severe attacks. We leave for future work the articulation of specific intervention strategies based on this model.

Because many questions remain about the accuracy of this model for understanding modern terrorism and its utility for counter-terrorism efforts, we remain modest about its long-term value. First, there is the question of the dependence of the central prediction---the power-law distribution in the frequency of severe attacks---on the particular assumptions we have described here. Already we have shown that the power-law prediction does not depend on the function $b(k)$, and it may be that other model assumptions can also be eliminated, relaxed or made more realistic while preserving this behavior (see, for instance, Ruszczycki et al.~\cite{ruszczycki:etal:2008}).

For example, in most conflicts, the number of radicalized individuals $N$ is unlikely to remain constant, and may not vary slowly relative to the replacement of individuals lost from counter-terrorism activities, etc., or relative to the aggregation-disintegration dynamics. Changes in $N$ should thus induce perturbations to the model's steady-state behavior. Further, empirical research may show that cells do not launch attacks with probability independent of their size. If larger cells launched attacks more frequently than smaller cells, it may be possible to adjust the aggregation dynamics so as to produce correspondingly fewer of these large cells, thus leaving the qualitative behavior of the model unchanged.

Existing analyses have focused on the steady-state behavior, but real organizations may exhibit a transient period of non-power-law behavior during which they self-organize to the critical state. The character and duration of this transient behavior depends on the initial distribution of cell sizes, but for reasonable initial conditions, it is unknown what specific behavior we should expect. Finally, the strategic utility of terrorist attacks is widely accepted~\cite{kydd:walter:2002,pape:2003,enders:sandler:2004,clauset:etal:2009:b}. However, the model assumes that attacks are largely stochastic in nature, and it is unknown whether these two perspectives can be reconciled. Research on this model would benefit greatly from mathematical generalizations that move us toward discovering the most general version that still produces the power-law distribution.

Second, although the model correctly predicts the distribution of event severities, this agreement is a relatively indirect test of the model's accuracy, and a stronger test would consider the accuracy of the model's specific assumptions or its predicted dynamics. Tests along these lines may also point out the most useful mathematical generalizations. For the record, we describe a number of ways the model can be tested.

Anecdotal evidence, including post-hoc analyses of severe events like the September 11th attacks~\cite{sageman:2004}, suggests that extremely severe attacks often require significantly more resources and manpower than small-severity attacks (see Clauset, Young, and Gleditsch~\cite{clauset:young:gleditsch:2007} for additional discussion), but it is unknown whether this is a systematic relationship and whether the precise form of assumption 5, i.e., $v(k)\propto k$, is sufficiently accurate. Without access to data on the internal dynamics of terrorist organizations, a direct test of this assumption seems impossible. However, research on determining what factors correlate with the severity of terrorist attacks may indirectly address this question (for instance, see Harrison~\cite{harrison:2006}, Clauset, Young, and Gleditsch~\cite{clauset:young:gleditsch:2007}, Asal and Rethemeyer~\cite{asal:rethemeyer:2008}, and Clauset and Gleditsch~\cite{clauset:gleditsch:2009}).

Further, the assumption that cells initiate attacks independently of their age or history may prove to be overly simplistic, and systematic correlations could produce deviations from the expected power-law form. That being said, recent work finds no significant deviations from a power-law distribution for attacks worldwide that killed at least 10 individuals~\cite{clauset:young:gleditsch:2007}, and it remains to be seen whether other kinds of systematic correlations exist. The model defined here also predicts that the severity of attacks by individual terrorist organizations should follow a power law. Johnson et al.~\cite{johnson:etal:2005,johnson:etal:2006} previously analyzed the conflict in Colombia, which is largely defined by the actions of the Revolutionary Armed Forces of Colombia (FARC) and found evidence supporting this fact. However, a more systematic study of individual organizations is needed to fully vet this hypothesis.

Deviations, however, may not mean that the entire model is incorrect. Non-power-law behavior could be indicative of the aforementioned transient, non-critical behavior. Additionally, the model assumes that terrorist cells only interact with other cells within the same organization, e.g., Taliban fighters do not aggregate with FARC fighters; however, some evidence suggests that cells sometimes do interact across organizational boundaries, for instance, between organizations involved in the same conflict~\cite{pedahzur:perliger:2006,araj:2008}, between allied organizations~\cite{sageman:2004}, or when fighters from different conflicts are jailed together~\cite{mckeown:2001}. Thus, in some cases, the set of cells that constitute a ``group'' in the sense of the model may not correspond to a single identifiable terrorist organization; instead, a ``group'' may be a somewhat amorphous set of cells, spread over multiple organizations. Thus, the set of events by which to test the power-law hypothesis may not always break cleanly at organizational boundaries. The extent to which the network of organizational alliances worldwide structures and constrains the set of possible interactions between cells is largely unknown, but likely plays an important role in the global dynamics of terrorism.

This discussion points to a more critical test of the accuracy of the model: validating the aggregation-disintegration dynamics themselves. As described above, without detailed data on the internal organizational dynamics or on the actions of many individual fighters, this part of the model seems difficult to test directly. Conventional wisdom suggests that aggregation-disintegration dynamics are unrealistic, as interactions between cells could pose security risks to the larger organization or to ongoing operations. However, recent analyses of organizations involved in the ``global jihad'' indicate that interactions, including aggregations, do indeed occur with some frequency, and that such interactions may be critical to the execution of particularly severe attacks~\cite{fouda:fielding:2003,sageman:2004}. But, it remains unclear how often such aggregations occur, how widespread they are, and how necessary they are to the execution of large attacks. Taking this anecdotal evidence at face value, it still remains unclear whether they occur frequently enough to allow an organization or set of cells to converge on the critical state---exhibiting the power-law distribution in cell sizes---in a timely fashion.

Ideally, all of these assumptions and predictions will be tested with empirical data to determine just how realistic, and thus how useful, this model is. Due to the scarcity of systematic, quantitative data on terrorism, some of these assumptions may prove impossible to test directly. On the other hand, by focusing on the model's testable predictions, it may be possible to test the model indirectly using available data. These empirical tests, along with the mathematical tests of the dependence of the power-law result on the model's particular assumptions, are promising avenues for future work on Richardson's Law as applied to the severity of terrorist events. \\

\begin{acknowledgments}
The authors thank Kristian S.\ Gleditsch, Christopher K.\ Butler, Libby Wood, Lars-Erik Cederman, Sidney Redner, and Neil F.\ Johnson for helpful conversations, and two anonymous referees for comments on an earlier draft of this manuscript. This work was supported in part by the Santa Fe Institute.
\end{acknowledgments}


\renewcommand{\thefigure}{A\arabic{figure}}
\setcounter{figure}{0}
\renewcommand{\thetable}{A\arabic{table}}
\setcounter{table}{0}

\begin{appendix}

\section{Power-law Distributions}
\label{appendix:powerlaws}
Some readers may be unfamiliar with power-law distributions (sometimes also called ``Zipf's law'' or ``Pareto distributions'' after two early researchers who championed their study~\cite{zipf:1949,pareto:1896}), and this appendix is to serve as a brief, and somewhat informal, primer on the topic. What distinguishes a power-law distribution from the more familiar Normal distribution is its {\em heavy tail}. That is, in a power law, there is a non-trivial amount of weight far from the distribution's center. This feature, in turn, implies that events orders of magnitude larger (or smaller) than the mean are relatively common. The latter point is particularly true when compared to a Normal distribution, where there is essentially no weight far from the mean. 

Although there are many distributions that exhibit heavy tails, the power law is special and exhibits a straight line with slope $\alpha$ on doubly-logarithmic axes. (Note that some data being straight on log-log axes is a necessary, but not a sufficient condition of being power-law distributed.) This behavior is termed \emph{scale invariance} because the power law admits the following property: multiplying its argument by some factor $k$ results in a change in the corresponding frequency that is independent of the function's argument. For example, if $p(x) = C x^{-\alpha}$, then
\begin{align}
p(k\cdot x) & = C \, k^{-\alpha} x^{-\alpha} \nonumber \\
 & = k^{-\alpha} \,p(x) \enspace, \nonumber
\end{align}
for every value $x$. For this reason, the exponent $\alpha$ is called the ``scaling exponent'' (for historical reasons, $\alpha-1$ is sometimes called the ``Pareto exponent''),  and the distribution is said to ``scale.'' This property also implies that there's no qualitative difference between large and small events.

Power-law distributed quantities are not uncommon, and many characterize the distribution of familiar quantities. For instance, consider the populations of the 600 largest cities in the United States (from the 2000 Census). Among these, the average population is only $\overline{x} =165~719$, and metropolises like New York City and Los Angles seem to be ``outliers'' relative to this size. One clue that city sizes are not well explained by a Normal distribution is that the sample standard deviation $\sigma = 410~730$ is significantly larger than the sample mean. Indeed, if we modeled the data in this way, we would expect to see $1.8$ times fewer cities at least as large as Albuquerque (population $448~607$) than we actually do. Further, because it is more than a dozen standard deviations above the mean, we would never expect to see a city as large as New York City (population $8~008~278$), and largest we expect would be Indianapolis (population $781~870$). 

\begin{figure}[t]
\begin{center}
\includegraphics[scale=0.48]{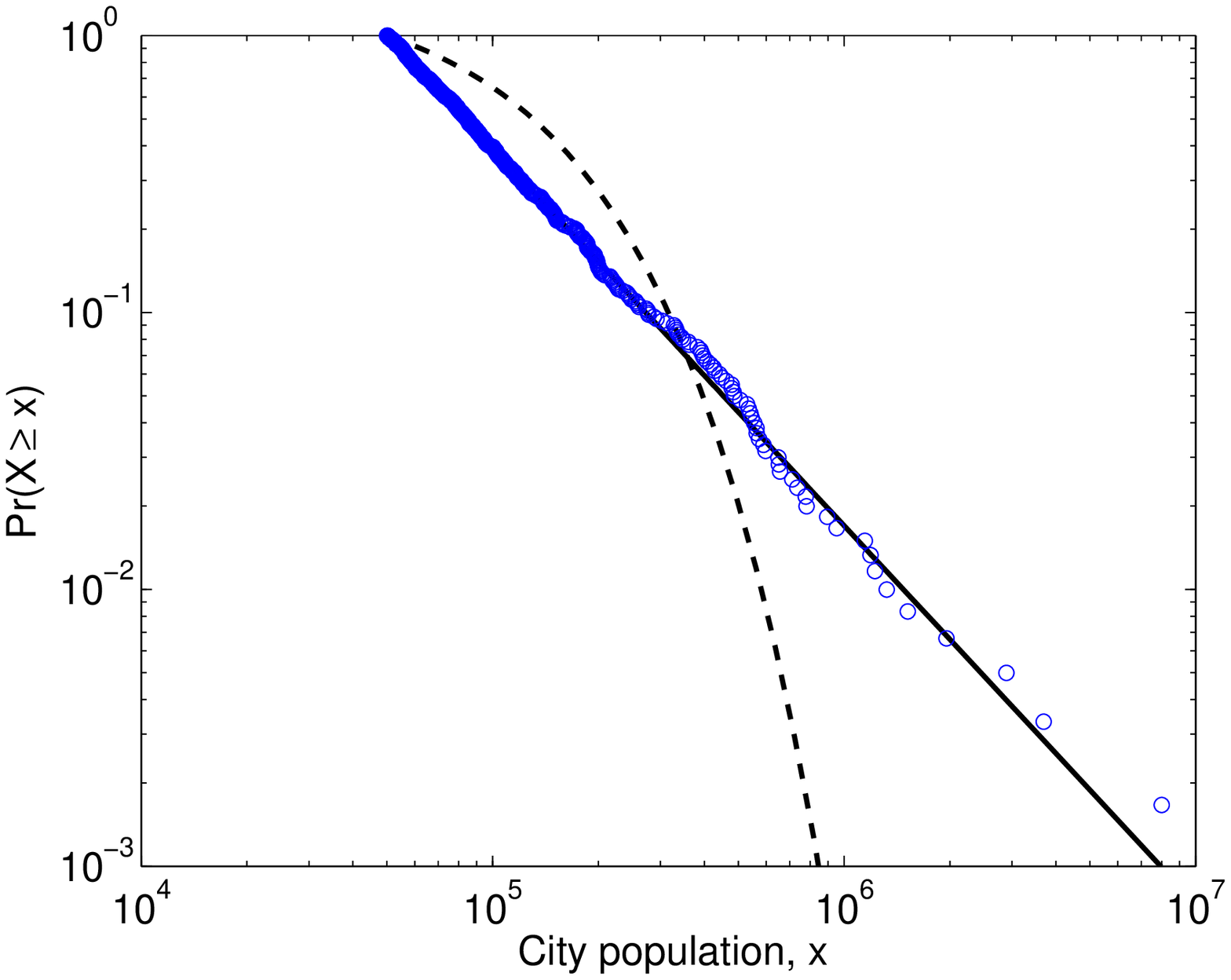}
\end{center}
\caption{The sizes of the 600 largest cities in the Unites States, i.e., those with population $x\geq50~000$, based on data from the 2000 Census. The data are plotted as a complementary cumulative distribution function $\Pr(X\geq x)$. The solid black line shows the power-law behavior that the distribution closely follows, with scaling exponent $\hat{\alpha}=2.36\pm0.06$, while the dashed black line shows a truncated normal distribution with the same sample mean.}
\label{fig:cities}
\end{figure}

Figure~\ref{fig:cities} shows the empirical data for these 600 cities, plotted on doubly-logarithmic axes as a complementary cumulative distribution function $\Pr(X\geq x)$ (the standard way of visualizing this kind of data). 
The scaling behavior of this empirical data is clear, and the corresponding power-law model (black line) a reasonably good fit. In contrast, the truncated normal model is a terrible fit. These notions of goodness-of-fit can be made precise using an appropriately defined significance test, such as the one described by Clauset, Shalizi, and Newman~\cite{clauset:etal:2009}.

As a more whimsical second example, consider a world where the heights of Americans were distributed as a power law, with approximately the same average as the true distribution (which is convincingly Normal when certain exogenous factors are controlled). In this case, we would expect nearly $60~000$ individuals to be as tall as the tallest adult male on record, at $2.72$ meters. Further, we would expect ridiculous facts such as $10~000$ individuals being as tall as an adult male giraffe, one individual as tall as the Empire State Building ($381$ meters), and $180$ million diminutive individuals standing a mere $17$ cm tall. In fact, this same analogy was recently used to describe the counter-intuitive nature of the extreme inequality in the wealth distribution in the United States~\cite{crook:2006}, whose upper tail is often said to follow a power law.

Although much more can be said about power laws, we hope that the curious reader takes away a few basic facts from this brief introduction. First, heavy-tailed distributions do not conform to our expectations of a linear, or normally distributed, world. As such, the average value of a power law is not representative of the entire distribution, and events orders of magnitude larger than the mean are, in fact, relatively common. Second, the scaling property of power laws implies that, at least statistically, there is no qualitative difference between small, medium and extremely large events, as they are all succinctly described by a very simple statistical relationship. Readers who would like more information about power laws should refer to the extensive reviews by Kleiber and Kotz~\cite{kleiber:kotz:2003}, Newman~\cite{newman:2005} and Mitzenmacher~\cite{mitzenmacher:2004:b}.

\section{Generating Functions}
\label{appendix:gfun}

Generating functions are a mathematical tool for representing and doing calculations with infinite sequences. Suppose you have two infinite sequences: $(c_{0},c_{1},c_{2},\dots)$ and $(d_{0},d_{1},d_{2},\dots)$. Their generating functions are defined by
\begin{align}
F(z) & \equiv \sum_{k=0}^{\infty}c_{k}z^{k} \enspace , \\
G(z) & \equiv \sum_{k=0}^{\infty}d_{k}z^{k} \enspace .
\end{align}
Both are analytic functions of the complex variable $z$. Their product \mbox{$H(z)=F(z)G(z)$} is a power series
\begin{align}
H(z) = & \sum_{k=0}^{\infty}h_{k}z^{k} 
\end{align}
with coefficients that are sums of products of the $c_{k}$ and $d_{k}$:
\begin{align}
h_{k} = & \sum_{m,n=0}^{\infty}\!\!{}^{'}c_{m}d_{n} \enspace ,
\end{align}
where again $\sum\!{}^{'}\,$ denotes a summation over all natural numbers $m$ and $n$ such that
$m+n=k$. This property was used in Section~\ref{sec:distribution}.

It is often easier to calculate a generating function than to work explicitly with the sequence of the expansion coefficients. Once the function is known explicitly, the coefficients can be calculated from Cauchy's theorem
\begin{align}
h_{k} = & \frac{1}{2\pi \imath} \oint_{C} H(z) \frac{\d z}{z^{k+1}} \enspace ,
\end{align}
where $C$ encircles the origin of the complex $z$-plane once in the counter-clockwise direction.

Readers who would like more information about generating functions and their use in mathematical analysis should refer to the textbook by Wilf~\cite{wilf:2006}.

\section{The cut-off $k_{0}$ and the value of $n_{1}^{\eq}$}
\label{appendix:cutoff}
The full equation for $n_{1}(t)$ follows from the model assumptions in Section~\ref{sec:model}. It has the form
\begin{align}
\frac{\d n_{1}}{\d t} & = \sum_{k=2}^{k_{0}} k\, b(k) n_{k} - A_{0}n_{1}\sum_{\ell=1}^{\infty}\ell^{a}n_{\ell} \enspace ,
\end{align}
which gives for the stationary state the equation
\begin{align}
\sum_{k=2}^{k_{0}} k\, b(k) n_{k}^{\eq} & = A_{0} n_{1}^{\eq} \sum_{\ell=1}^{\infty} \ell^{a}n_{\ell}^{\eq} \enspace .
\end{align}
This equation connects the cut-off $k_{0}$ with $n_{1}^{\eq}$. The right-hand side equals $A_{0}n_{1}^{\eq}f(1)$, where Eq.~\eqref{eq:3.4} was used. Using Eq.~\eqref{eq:15}, one can rewrite this as
\begin{align}
\label{eq:c:3}
\sum_{k=2}^{k_{0}}k\, b(k) n_{k}^{\eq} & = 2 A_{0}(n_{1}^{\eq})^{2} \enspace .
\end{align}

The value of $n_{1}^{\eq}$ can then be calculated from the relation
\begin{align}
\label{eq:c:4}
N & = n_{1}^{\eq} + \sum_{k=2}^{\infty} k n_{k}^{\eq} \enspace ,
\end{align}
which expresses the fact that the total number of radicalized individuals should equal $N$. For the case $a=1$, the definition in Eq.~\eqref{eq:3.4} shows that one can rewrite Eq.~\eqref{eq:c:4} in the form
\begin{align}
\sum_{k=1}^{\infty} k\, n_{k}^{\eq} & = N = f(1) \enspace .
\end{align}
Combining this expression with Eq.~\eqref{eq:15} gives $N=2n_{1}^{\eq}$, so one finds
\begin{align}
\label{eq:c:6}
n_{1}^{\eq} & = \frac{1}{2}N\enspace ,
\end{align}
that is: half the number of these individuals are singletons and half that number are part of larger cells.

To now calculate the cut-off $k_{0}$ (for $k>k_{0}$ we assume $n_{k}^{\eq}=0$), one rewrites Eq.~\eqref{eq:c:3} in the form
\begin{align}
\label{eq:c:7}
\sum_{k=2}^{k_{0}} k\, b(k) n_{k}^{\eq} & = \frac{1}{2}A_{0}N^{2} \enspace .
\end{align}
As an example for explicit calculation, we take the case $a=1$ and
\begin{align}
b(k) & = B_{0}k^{b}\enspace,
\end{align}
for $\frac{1}{2} < b < \frac{3}{2}$, where the exponent $b$ is some number in the vicinity of unity. Equation~\eqref{eq:26} now gives Eq.~\eqref{eq:c:7} the form
\begin{align}
\label{eq:c:9}
\sum_{k=2}^{k_{0}} k\, b(k) n_{k}^{\eq} & \cong \frac{B_{0}N}{2\sqrt{\pi}}\sum_{k=2}^{k_{0}} k^{b-3/2}\enspace ,
\end{align}
where a small error is neglected, which is due to the fact that we used the $k\gg1$ asymptotic expression for $n_{k}^{\eq}$ for all $k\geq2$. The series in the right-hand of Eq.~\eqref{eq:c:9} can be approximated by an integral, which gives
\begin{align}
\sum_{k=2}^{k_{0}} k^{b-3/2} & \cong \int_{2}^{k_{0}} k^{b-3/2} \d k \nonumber \\
& \cong \left(\frac{1}{b-\frac{1}{2}}\right)k_{0}^{b-1/2} \enspace ,
\end{align}
for $k_{0}\gg1$. With these results, Eq.~\eqref{eq:c:7} takes the form
\begin{align}
\frac{B_{0}}{2\sqrt{\pi}}\left(\frac{1}{b-\frac{1}{2}}\right)k_{0}^{b-1/2} & = \frac{1}{2}A_{0}N \enspace ,
\end{align}
which gives an explicit value for the cut-off:
\begin{align}
\label{eq:c:12}
k_{0} & = \left[ \frac{A_{0}}{B_{0}}\left(b-\frac{1}{2}\right)\sqrt{\pi}N \right]^{1\left/\left(b-\frac{1}{2}\right)\right.} \enspace .
\end{align}
The essential feature of this result is that $k_{0}\gg1$ when $N\gg1$. At the cut-off, the value of $n_{k_{0}}^{\eq}$ is proportional to a negative power of $N$:
\begin{align}
n_{k_{0}}^{\eq} & \propto N^{1-\frac{5}{2}\left(b-\frac{1}{2}\right)^{-1}} \enspace ,
\end{align}
where one uses Eqs.~\eqref{eq:26},~\eqref{eq:c:6} and~\eqref{eq:c:12}. Hence for $k>k_{0}$, all numbers $n_{k}^{\eq} \ll1$ and are therefore irrelevant. These features of the cut-off show that its existence is a mathematical artifact only, with no consequences for the distribution of cell sizes for realistic values of $k$.

\end{appendix}




\begin{thebibliography}{}

\bibitem[\protect\citeauthoryear{Araj}{Araj}{2008}]{araj:2008}
Araj, B. (2008).
\newblock Harsh state repression as a cause of suicide bombing: {T}he case of
  the {P}alestinian-{I}sraeli conflict.
\newblock {\em Studies in Conflict \& Terrorism\/}~{\em 31\/}(4), 284--303.

\bibitem[\protect\citeauthoryear{Asal and Rethemeyer}{Asal and
  Rethemeyer}{2008}]{asal:rethemeyer:2008}
Asal, V. and R.~K. Rethemeyer (2008).
\newblock The nature of the beast: {O}rganizational structures and the
  lethality of terrorist attacks.
\newblock {\em Journal of Politics\/}~{\em 70\/}(2), 437--449.

\bibitem[\protect\citeauthoryear{Bak, Tang, and Wiesenfeld}{Bak
  et~al.}{1987}]{bak:tang:wiesenfeld:1987}
Bak, P., C.~Tang, and K.~Wiesenfeld (1987).
\newblock Self-organized criticality: An explanation of 1/f noise.
\newblock {\em Physical Review Letters\/}~{\em 59\/}(4), 381--384.

\bibitem[\protect\citeauthoryear{Biggs}{Biggs}{2005}]{biggs:2005}
Biggs, M. (2005).
\newblock Strikes as forest fires: {C}hicago and {P}aris in the late 19th
  century.
\newblock {\em American Journal of Sociology\/}~{\em 111\/}(1), 1684--1714.

\bibitem[\protect\citeauthoryear{Cederman}{Cederman}{2003}]{cederman:2003}
Cederman, L.-E. (2003).
\newblock Modeling the size of wars: From billiard balls to sandpiles.
\newblock {\em American Political Science Review\/}~{\em 97\/}(1), 135--150.

\bibitem[\protect\citeauthoryear{Cioffi-Revilla and Romero}{Cioffi-Revilla and
  Romero}{2009}]{cioffi-revilla:romero:2009}
Cioffi-Revilla, C. and P.~P. Romero (2009).
\newblock Modeling uncertainty in adversary behavior: {A}ttacks in {D}iyala
  province, {I}raq, 2002--2006.
\newblock {\em Studies in Conflict \& Terrorism\/}~{\em 32\/}(3), 253--276.

\bibitem[\protect\citeauthoryear{Clauset and Gleditsch}{Clauset and
  Gleditsch}{2009}]{clauset:gleditsch:2009}
Clauset, A. and K.~S. Gleditsch (2009).
\newblock Developmental dynamics of terrorist organizations.
\newblock Preprint, {\tt http://arxiv.org/abs/0906.3287} (accessed June 17,
  2009).

\bibitem[\protect\citeauthoryear{Clauset, Heger, Young, and Gleditsch}{Clauset
  et~al.}{ming}]{clauset:etal:2009:b}
Clauset, A., L.~Heger, M.~Young, and K.~S. Gleditsch ({Forthcoming}).
\newblock The strategic calculus of terrorism: {S}ubstitution and competition
  in the {I}srael-{P}alestine conflict.
\newblock {\em Cooperation \& Conflict}.

\bibitem[\protect\citeauthoryear{Clauset, Shalizi, and Newman}{Clauset
  et~al.}{2009}]{clauset:etal:2009}
Clauset, A., C.~R. Shalizi, and M.~E.~J. Newman (2009).
\newblock Power-law distributions in empirical data.
\newblock {\em SIAM Review\/}~{\em 51}, 661--703.

\bibitem[\protect\citeauthoryear{Clauset, Young, and Gleditsch}{Clauset
  et~al.}{2007}]{clauset:young:gleditsch:2007}
Clauset, A., M.~Young, and K.~S. Gleditsch (2007).
\newblock On the frequency of severe terrorist events.
\newblock {\em Journal of Conflict Resolution\/}~{\em 51\/}(1), 58--87.

\bibitem[\protect\citeauthoryear{Cordes, Jenkins, Kellen, Bass, Relles, Sater,
  Juncosa, Fowler, and Petty}{Cordes et~al.}{1985}]{cordes:etal:1985}
Cordes, B., B.~M. Jenkins, K.~Kellen, G.~Bass, D.~Relles, W.~Sater, M.~Juncosa,
  W.~Fowler, and G.~Petty (1985).
\newblock {\em A Conceptual Framework for Analyzing Terrorist Groups}.
\newblock Arlington: RAND Corporation.

\bibitem[\protect\citeauthoryear{Cronin}{Cronin}{2003}]{cronin:2003}
Cronin, A.~K. (2003).
\newblock Behind the curve.
\newblock {\em International Security\/}~{\em 27\/}(3), 30--58.

\bibitem[\protect\citeauthoryear{Crook}{Crook}{2006}]{crook:2006}
Crook, C. (2006).
\newblock The height of inequality.
\newblock {\em The Atlantic Monthly\/}~{\em 298\/}(2), 36--37.

\bibitem[\protect\citeauthoryear{Enders and Sandler}{Enders and
  Sandler}{2000}]{enders:sandler:2000}
Enders, W. and T.~Sandler (2000).
\newblock Is transnational terrorism becoming more threatening? {A} time-seires
  investigation.
\newblock {\em Journal of Conflict Resolution\/}~{\em 44\/}(3), 307--332.

\bibitem[\protect\citeauthoryear{Enders and Sandler}{Enders and
  Sandler}{2004}]{enders:sandler:2004}
Enders, W. and T.~Sandler (2004).
\newblock What do we know about the substitution effect in transnational
  terrorism?
\newblock In A.~Silke (Ed.), {\em Researching Terrorism Trends, Achievements,
  Failures}. London: Frank Cass.

\bibitem[\protect\citeauthoryear{Fouda and Fielding}{Fouda and
  Fielding}{2003}]{fouda:fielding:2003}
Fouda, Y. and N.~Fielding (2003).
\newblock {\em Masterminds of Terror: {T}he Truth Behind the Most Devastating
  Terrorist Attack the World Has Ever Seen}.
\newblock New York: Arcade.

\bibitem[\protect\citeauthoryear{Harrison}{Harrison}{2006}]{harrison:2006}
Harrison, M. (2006).
\newblock Bombers and bystanders in suicide attacks in {I}srael.
\newblock {\em Studies in Conflict \& Terrorism\/}~{\em 29}, 187--206.

\bibitem[\protect\citeauthoryear{Johnson, Spagat, Restrepo, Bohorquez, Suarez,
  Restrepo, and Zarama}{Johnson et~al.}{2005}]{johnson:etal:2005}
Johnson, N.~F., M.~Spagat, J.~Restrepo, J.~Bohorquez, N.~Suarez, E.~Restrepo,
  and R.~Zarama (2005).
\newblock From old wars to new wars and global terrorism.
\newblock Preprint. {\tt http://arxiv.org/physics/0506213} (accessed on June
  29, 2005).

\bibitem[\protect\citeauthoryear{Johnson, Spagat, Restrepo, Becerra, Bohorquez,
  Suarez, Restrepo, and Zarama}{Johnson et~al.}{2006}]{johnson:etal:2006}
Johnson, N.~F., M.~Spagat, J.~A. Restrepo, O.~Becerra, J.~C. Bohorquez,
  N.~Suarez, E.~M. Restrepo, and R.~Zarama (2006).
\newblock Universal patterns underlying ongoing wars and terrorism.
\newblock Preprint. {\tt http://arxiv.org/physics/0605035} (accessed on May 3,
  2006).

\bibitem[\protect\citeauthoryear{Kleiber and Kotz}{Kleiber and
  Kotz}{2003}]{kleiber:kotz:2003}
Kleiber, C. and S.~Kotz (2003).
\newblock {\em Statistical Size Distributions in Economics and Actuarial
  Sciences}.
\newblock Hoboken, NJ: Wiley.

\bibitem[\protect\citeauthoryear{Kydd and Walter}{Kydd and
  Walter}{2002}]{kydd:walter:2002}
Kydd, A. and B.~Walter (2002).
\newblock Sabotaging the peace: The politics of extremist violence.
\newblock {\em International Organization\/}~{\em 56\/}(2), 263--296.

\bibitem[\protect\citeauthoryear{Lacina}{Lacina}{2006}]{lacina:2006}
Lacina, B. (2006).
\newblock Explaining the severity of civil wars.
\newblock {\em Journal of Conflict Resolution\/}~{\em 50\/}(2), 276--289.

\bibitem[\protect\citeauthoryear{Li}{Li}{2005}]{li:2005}
Li, Q. (2005).
\newblock Does democracy promote or reduce transnational terrorist incidents?
\newblock {\em Journal of Conflict Resolution\/}~{\em 49\/}(2), 278--297.

\bibitem[\protect\citeauthoryear{Malamud, Morein, and Turcotte}{Malamud
  et~al.}{1998}]{malamud:etal:1998}
Malamud, B.~D., G.~Morein, and D.~L. Turcotte (1998).
\newblock Forest fires: {An} example of self-organized critical behavior.
\newblock {\em Science\/}~{\em 281}, 1840--1841.

\bibitem[\protect\citeauthoryear{McKeown}{McKeown}{2001}]{mckeown:2001}
McKeown, L. (2001).
\newblock {\em Out of Time: {I}rish {R}epublican Prisoners {L}ong {K}esh
  1972-2000}.
\newblock Belfast, UK: Beyond the Pale Publications.

\bibitem[\protect\citeauthoryear{Mitzenmacher}{Mitzenmacher}{2004}]{mitzenmach%
er:2004:b}
Mitzenmacher, M. (2004).
\newblock A brief history of generative models for power law and lognormal
  distributions.
\newblock {\em Internet Mathematics\/}~{\em 1\/}(2), 226--251.

\bibitem[\protect\citeauthoryear{{National Memorial Institute for the
  Prevention of Terrorism}}{{National Memorial Institute for the Prevention of
  Terrorism}}{2008}]{mipt:2008}
{National Memorial Institute for the Prevention of Terrorism} (2008).
\newblock Terrorism {K}nowledge {B}ase.
\newblock {\tt http://www.tkb.org} (access date January 29, 2008).

\bibitem[\protect\citeauthoryear{Newman}{Newman}{2005}]{newman:2005}
Newman, M. E.~J. (2005).
\newblock Power laws, {P}areto distributions and {Z}ipf's law.
\newblock {\em Contemporary Physics\/}~{\em 46\/}(5), 323--351.

\bibitem[\protect\citeauthoryear{Pape}{Pape}{2003}]{pape:2003}
Pape, R.~A. (2003).
\newblock The strategic logic of suicide terrorism.
\newblock {\em American Political Science Review\/}~{\em 97\/}(3), 343--361.

\bibitem[\protect\citeauthoryear{Pareto}{Pareto}{1896}]{pareto:1896}
Pareto, V. (1896).
\newblock {\em Cours d'{E}conomie Politique}.
\newblock Geneva: Droz.

\bibitem[\protect\citeauthoryear{Pedahzur and Perliger}{Pedahzur and
  Perliger}{2006}]{pedahzur:perliger:2006}
Pedahzur, A. and A.~Perliger (2006).
\newblock The changing nature of suicide attacks: {A} social network
  perspective.
\newblock {\em Social Forces\/}~{\em 84\/}(4), 1987--2008.

\bibitem[\protect\citeauthoryear{Richardson}{Richardson}{1948}]{richardson:194%
8}
Richardson, L.~F. (1948).
\newblock Variation of the frequency of fatal quarrels with magnitude.
\newblock {\em Journal of the American Statistical Association\/}~{\em 43},
  523--546.

\bibitem[\protect\citeauthoryear{Richardson}{Richardson}{1960}]{richardson:196%
0}
Richardson, L.~F. (1960).
\newblock {\em Statistics of Deadly Quarrels}.
\newblock Pittsburgh: The Boxwood Press.

\bibitem[\protect\citeauthoryear{Ruszczycki, Burnett, Zhao, and
  Johnson}{Ruszczycki et~al.}{2008}]{ruszczycki:etal:2008}
Ruszczycki, B., B.~Burnett, Z.~Zhao, and N.~F. Johnson (2008).
\newblock Relating the microscopic rules in coalescence-fragmentation models to
  the emergent cluster-size distribution.
\newblock Preprint. {\tt http://arxiv.org/abs/0808.0032} (accessed on June 8,
  2009).

\bibitem[\protect\citeauthoryear{Sageman}{Sageman}{2004}]{sageman:2004}
Sageman, M. (2004).
\newblock {\em Understanding Terror Networks}.
\newblock Philadelpha: University of Pennsylvania Press.

\bibitem[\protect\citeauthoryear{Simon}{Simon}{1955}]{simon:1955}
Simon, H.~A. (1955).
\newblock On a class of skew distribution functions.
\newblock {\em Biometrika\/}~{\em 42\/}(34), 425--440.

\bibitem[\protect\citeauthoryear{Small and Singer}{Small and
  Singer}{1982}]{small:singer:1982}
Small, M. and J.~D. Singer (1982).
\newblock {\em Resort to Arms: {I}nternational and Civil Wars, 1816--1980}.
\newblock Beverly Hills: Sage Publications.

\bibitem[\protect\citeauthoryear{{United States Department of State}}{{United
  States Department of State}}{2004}]{state:2004}
{United States Department of State} (2004).
\newblock Patterns of {G}lobal {T}errorism, 2003.
\newblock Washington, DC, U.S. Department of State, April. {\tt
  http://www.state.gov/documents/organization/31912.pdf} (accessed on January
  22, 2005).

\bibitem[\protect\citeauthoryear{Wilf}{Wilf}{2006}]{wilf:2006}
Wilf, H. (2006).
\newblock {\em generatingfunctionology\/} (3 ed.).
\newblock A. K. Peters Ltd.

\bibitem[\protect\citeauthoryear{Zipf}{Zipf}{1949}]{zipf:1949}
Zipf, G.~K. (1949).
\newblock {\em Human Behavior and the principle of least effort}.
\newblock Cambridge, MA: Addison-Wesley.

\end{thebibliography}

\end{document}